\documentclass[10pt,a4paper,twocolumn,tightenlines,amsmath,amssymb,nofootinbib,superscriptaddress,showpacs,accepted=2017-07-11]{quantumarticle}
\pdfoutput=1
\usepackage{graphicx}
\usepackage{dcolumn}
\usepackage{bm}
\usepackage{latexsym,epsfig,epstopdf}
\usepackage{float}
\usepackage{amsmath,amssymb,amsthm}
\usepackage{xspace}
\usepackage{graphicx}
\usepackage{latexsym}
\usepackage[mathscr]{eucal}
\usepackage{dcolumn}
\usepackage{array}
\usepackage[mathlines]{lineno}
\usepackage[numbers]{natbib}

\newtheorem*{rep@lemma}{\rep@title}
\newcommand{\newreplemma}[2]{%
\newenvironment{rep#1}[1]{%
 \def\rep@title{#2 \ref{##1}}%
 \begin{rep@lemma}}%
 {\end{rep@lemma}}}
\makeatother

\newtheorem{lemma}{Lemma}
\newreplemma{lemma}{Lemma}

\newtheorem*{result}{Central Result}

\newtheorem{definition}{Definition}
\newtheorem*{definition*}{Definition}
\newtheorem*{theorem*}{Central Result}
\newtheorem*{lemma*}{Lemma}
\newtheorem*{corollary*}{Corollary}

\makeatletter
\DeclareRobustCommand{\cev}[1]{%
  \mathpalette\do@cev{#1}%
}
\newcommand{\do@cev}[2]{%
  \fix@cev{#1}{+}%
  \reflectbox{$\m@th#1\vec{\reflectbox{$\fix@cev{#1}{-}\m@th#1#2\fix@cev{#1}{+}$}}$}%
  \fix@cev{#1}{-}%
}
\newcommand{\fix@cev}[2]{%
  \ifx#1\displaystyle
    \mkern#23mu
  \else
    \ifx#1\textstyle
      \mkern#23mu
    \else
      \ifx#1\scriptstyle
        \mkern#22mu
      \else
        \mkern#22mu
      \fi
    \fi
  \fi
}

\makeatletter
\newcommand*{\balancecolsandclearpage}{%
  \close@column@grid
  \clearpage
  \twocolumngrid
}
\makeatother

\newcommand{\ket}[1]{\left | #1 \right\rangle}
\newcommand{\bra}[1]{\left \langle #1 \right |}

\newcommand{\Tr}{\mathrm{Tr}}
\newcommand{\braket}[2]{\left\langle #1|#2\right\rangle}
\newcommand{\ketbra}[2]{|#1\left\rangle\right\langle #2 |}

\newcommand{\past}[1]{\cev{#1}}
\newcommand{\future}[1]{\vec{#1}}
\newcommand{\lx}[0]{\past{x}}
\newcommand{\rx}[0]{\future{x}}
\newcommand{\Lx}[0]{\past{X}}
\newcommand{\Rx}[0]{\future{X}}

\DeclareMathOperator{\tr}{Tr}

\providecommand{\Pr}{}
\renewcommand{\Pr}[1]{\mathrm{P}\!\left(#1\right)}
\newcommand{\cPr}[2]{\mathrm{P}\!\left(#1 \,|\, #2\right)}

\newcommand{\Ent}[1]{{H}\hspace{-0.25em}\left(#1\right)}

\newcommand{\cPrline}[2]{\mathrm{P}(#1 \,|\, #2)}
\newcommand{\Prline}[1]{\mathrm{P}(#1)}

\newcommand{\inlineheading}[1]{\textbf{{#1}}}

\newcommand{\CQT}{Centre for Quantum~Technologies, National~University~of~Singapore, 3 Science Drive 2, 117543, Singapore}

\newcommand{\Oxf}{Atomic~and~Laser~Physics, University~of~Oxford, Clarendon~Laboratory, Parks Road, Oxford, OX1 3PU, United Kingdom.}
\newcommand{\NTU}{School of Mathematical and Physical Scieces, Nanyang Technological University, Singapore}
\newcommand{\complexity}{Complexity Institute, Nanyang Technological University, Singapore}

\begin{document}

\title{The classical--quantum divergence of complexity in modelling spin chains}

\author{Whei Yeap Suen}
\email{wheiyeap@u.nus.edu}
\affiliation{\CQT}

\author{Jayne Thompson}
\affiliation{\CQT}

\author{Andrew J.\ P.\ Garner}
\affiliation{\CQT}

\author{Vlatko Vedral}
\affiliation{\Oxf}
\affiliation{\CQT}
\affiliation{Department of Physics, National University of Singapore, 3 Science Drive 2, Singapore 117543}

\author{Mile Gu}
\email{mgu@quantumcomplexity.org}
\affiliation{\NTU}
\affiliation{\complexity}
\affiliation{\CQT}

\date{\today}

\begin{abstract}
The minimal memory required to model a given stochastic process---known as the statistical complexity---is a widely adopted quantifier of structure in complexity science.
Here, we ask if quantum mechanics can fundamentally change the qualitative behaviour of this measure. We study this question in the context of the classical Ising spin chain.
In this system, the statistical complexity is known to grow monotonically with temperature.
We evaluate the spin chain's quantum mechanical statistical complexity by explicitly constructing its provably simplest quantum model, and demonstrate that this measure exhibits drastically different behaviour:
 it rises to a maximum at some finite temperature then tends back towards zero for higher temperatures.
This demonstrates how complexity, as captured by the amount of memory required to model a process, can exhibit radically different behaviour when quantum processing is \mbox{allowed}.
\end{abstract}


\maketitle

{\em Statistical complexity} emerges from the scientific ideal that we understand reality through cause and effect -- the more precisely we can isolate the causes of natural things, the greater our comprehension. In this context, the more causes one must postulate to fully replicate the behaviour of a process, the more complicated that process appears.  This motivates the statistical complexity as a measure of the process's intrinsic structure and complexity, since it represents the minimal amount of causal information one must record about past observations of a phenomena to model the statistics of observations made at future times~\cite{crutchfield1989inferring}.

Following these motivations, there has been significant work on developing a complete framework for statistical complexity. There are methods for evaluating the statistical complexity of general stationary stochastic processes~\cite{crutchfield1994calculi,shalizi2001computational}, and for estimating its value directly from observational statistics~\cite{shalizi2001causal, shalizi2004blind,kelly2012new}. Meanwhile, the corresponding optimal models can be systematically constructed. If a process has a statistical complexity of $C$, we can systematically replicate the process's statistical behaviour using a model that records only $C$ bits of information about the past. This analytic tractability, combined with a clear operational motivation, has propelled the use of statistical complexity within complexity science as a key measure of structure. Its field of study -- computational mechanics -- has been applied to analyse structure in diverse settings~\cite{perry1999finite, hanson1997computational, gonccalves1998inferring, tino1999extracting}, an early example being the Ising spin chain~\cite{crutchfield1997statistical}.

Conventional studies only consider building classical models. Yet, recent advances show that even when modelling the same classical process, a quantum model may require less input information~\cite{gu2012quantum,tan2014towards,mahoney2016occam,riechers2015closed,palsson2016experimental}. This motivates an important question: could statistical complexity exhibit very different qualitative behaviour in the quantum regime? If true, this would imply that many existing studies in statistical complexity could draw very different conclusions when taking quantum information processing into account. Progress in this direction, however, has faced significant hurdles. While there are constructions for quantum models that use less memory than the classical limit, it is not known whether these quantum models are optimal. Thus the {\em quantum statistical complexity} has never been explicitly evaluated for any process where there is a quantum advantage, and existing quantum models provide only upper bounds~\cite{tan2014towards}.

Here, we present the first provably optimal quantum model for a process with a quantum advantage -- a model of sequential measurement outcomes along a classical Ising spin chain. We determine the Ising chain's exact quantum statistical complexity. We compare this to existing studies of the spin chain's classical statistical complexity~\cite{crutchfield1997statistical}. Whereas the classical statistical complexity monotonically increases with temperature, the quantum measure attains a maximal value at some finite temperature, then decreases monotonically to zero. This demonstrates that the quantum and classical statistical complexities can diverge significantly in their qualitative behaviour. A system can become increasingly more difficult to model classically as certain physical parameters are varied, and yet simpler to model quantum mechanically.

\textbf{Framework.} Consider a stochastic system that at each discrete time $t \in \mathbb{Z}$ emits an output $x_t$ from some configuration space $\mathcal{X}$. The statistical behaviour at each time $t$ is characterised by a random variable $X_t$.
At each time $t$, the resulting string of outputs may be partitioned into a {\em past} $\Lx_t := \ldots X_{t-1} X_t$ comprised of outputs already emitted,  and a {\em  future} $\Rx_t := X_{t+1} X_{t+2} \ldots$ consisting of outputs yet to come.
The system's output behaviour can be captured by a stochastic process -- a probability distribution over bi-infinite strings $\Prline{\past{X}_t, \future{X}_t}$ that describes how past observations correlate with future behaviour. Here we consider {\em stationary} processes, where $P(\future{X}_t)= P(\future{X}_0)$ for any $t \in \mathbb{Z}^+$, and so hereafter drop the subscript $t$ from infinite strings.

Each instance of the process will have some specific past $\past{x}$, with a corresponding conditional future distribution $\cPrline{\future{X}}{\past{X}=\past{x}}$. For any non-trivial process, $\cPrline{\future{X}}{\past{X}=\past{x}}$ depends on $\past{x}$ -- this reflects the ideal that prediction is meaningful only when the past is correlated with the future. A predictive model, then, is an algorithmic abstraction of $\cPrline{\future{X}}{\past{X}}$ that exhibits statistically identical conditional future behaviour. Each model specifies an encoding function $\epsilon$ that maps each possible past $\past{x}$ onto an internal state $s \in \mathcal{S}$ of some physical system, such that systematic actions on this system at future time-steps will generate a string of outputs obeying $\cPrline{\future{X}}{\past{X}=\past{x}}$.

The stochastic process can then be modeled by Markovian dynamics on $\mathcal{S}$.
At each time-step $t$, a model in state $s_t$ generates $x$ according to $P(X_{t+1}=x|s_t) = P(X_{t+1}=x|\lx_t)$, and updates its internal state to $s_{t+1} = \epsilon(\past{x}')$, where $\past{x}' = \ldots x_{t-1}x_tx$ is the updated past.
This implies that if we have two black boxes, one containing an instance of the original process with specific past $\lx$, and another containing a machine initialized in state $s =  \epsilon(\past{x})$, then there will be no way to discriminate between them based on their observed output behaviour.
In \mbox{particular}, $\cPrline{\future{X}}{\past{X}\!=\!\past{x}} = \cPrline{\future{X}}{S\!=\!s}$, where $S$ is the random variable governing $s = \epsilon(\past{x})$.
The resulting models satisfy \emph{unifilarity}, such that there is no uncertainty in $S_{t+1}$ when given $X_{t+1}$ and $S_t$ -- an important mathematical property that ensures a model's internal state contains no information about future outputs outside of what is available from the past~\cite{crutch_sync}.
The amount of information such a model tracks is then given by the Shannon entropy, $\Ent{S} = -\sum_{s\in\mathcal{S}} p_s \log p_s$, where $p_s$ is the probability $\epsilon(\past{x}) = s$.

Clearly many different predictive models exist for each stochastic process.
The brute force approach would be to take $\epsilon$ as the identity map, yielding a model that demands the entire past as input.
While such a model will no doubt work, it is clearly inefficient.
Take for example, the modelling of a completely random process.
Here, as all pasts are equally likely, such a model would demand an unbounded amount of past information.
On the other hand, if the process is completely random, the future does not depend on the past, and thus no past information should be require to correctly sample the process's conditional future.

\begin{figure}[tbh]
\centering
\includegraphics[width = 0.4 \textwidth]{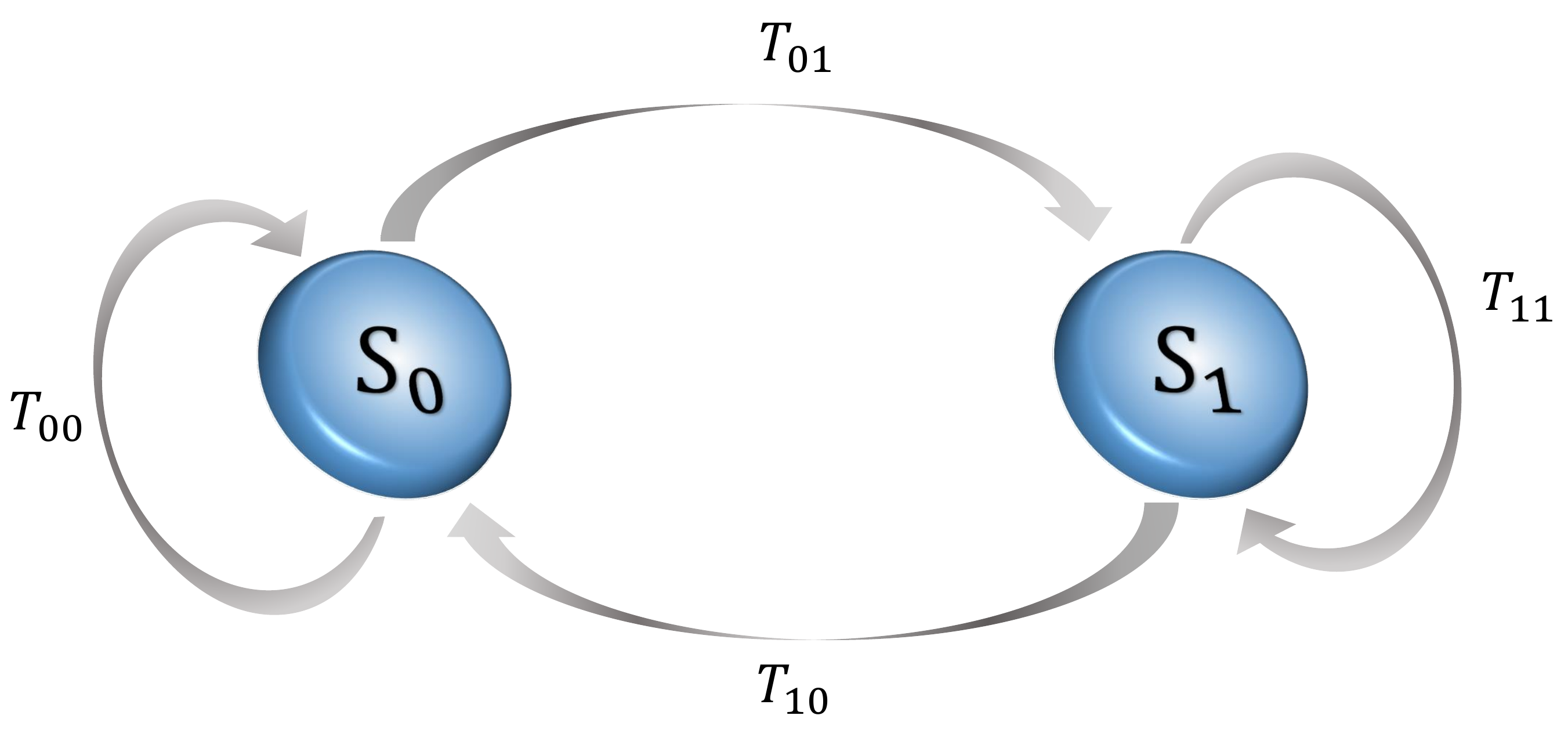}
\caption{The $\epsilon$-machine for the Ising chain consists of two causal states $s_0 = \{\lx, x_0 = 1\}$, and  $s_1 = \{\lx, x_0 = -1\}$. For this particular process, the $\epsilon$-machine has the special property such that $T^{(r)}_{ij} = 0$ for all $r \neq j$, and thus its dynamics can be entirely captured by transition probabilities $T_{ij} = T^{j}_{ij}$, the probability a machine in state $s_i$ will transition to state $s_j$ and output $j$, and coincides with the probability $P(X_{t+1} = j|X_{t} = i)$. The exact values for $T_{ij}$ depend on $B$, $J$ and $T$, and can be found in~\cite{feldman1998computational, yeomans1992statistical}.}
\label{emachine}
\end{figure}

\textbf{Simplest Classical Models.} Computational mechanics provides a framework for constructing the provably simplest predictive models of a process - the ones for which $\Ent{S}$ is minimized \cite{crutchfield1989inferring,shalizi2001computational}. This optimality can be achieved by introducing the equivalence relation
\begin{eqnarray}
\past{x} \sim_\epsilon \past{x}'
\text{ iff }
\cPr{\future{X}}{ \past{X}=\past{x} }
=
\cPr{ \future{X}} {\past{X}=\past{x}'} \label{eqclass}.
\end{eqnarray}
Let $\mathcal{S} = \{s\}$ be the resulting set of equivalence classes on the set of all pasts, and the encoding function be defined such that $\epsilon(\past{x}) = s$ iff $\past{x} \in s$. The resulting map represents a model that stores not which past was observed, but rather only which equivalence class that past belongs in. Given $s = \epsilon(\past{x})$, it is possible to systematically sample $\cPrline{\future{X}}{\past{X}=\past{x}}$. Specifically, consider a Markov finite state machine with state space $\mathcal{S}$. At each time-step, the machine's behaviour is dictated by transition elements $T^r_{ij} = \cPrline{S_{t+1}\!=\!s_j, X_{t+1}\!=\!r}{S_t\!=\!s_i}$: the probability that a machine initially in state $s_i$ will transition to $s_j$ whilst outputting $r\in\mathcal{X}$ (e.g.\ figure~\ref{emachine}). This ensures that when initialized in the appropriate causal state $\epsilon(\past{x})$, the machine can sequentially generate an arbitrarily long string $x_{t+1}x_{t+2} \ldots$ with probability $\cPrline{X_{t+1}X_{t+2} \ldots}{\past{x}}$.

In literature, such machines are referred to as $\epsilon$-machines. Meanwhile, the states $s \in \mathcal{S}$ are named \emph{causal states}, emphasizing that they capture all the causal information contained in $P(\past{X},\future{X})$~\cite{crutchfield1989inferring,shalizi2001computational}. The amount of information required to store these causal states
\begin{equation}
C_\mu = - \sum_{s\in\mathcal{S}} p_s \log p_s
\end{equation}
thus defines the \emph{statistical complexity} of a stochastic process -- representing the minimal amount of past information one needs about the process to replicate its conditional future statistics.


\inlineheading{Spin Chains.} We consider the 1D Ising system with nearest-neighbour interactions (as in~\cite{crutchfield1997statistical}) of length \mbox{$2N+1$} with periodic boundary conditions. This system is described by the Hamiltonian~\cite{yeomans1992statistical}
\begin{eqnarray}
\mathcal{H}_N &=&\sum_{k = -N \dots N} {-Jx_k x_{k+1} - Bx_k}, \label{hamiltonian}
\end{eqnarray}
where $J$ is the coupling parameter, $B$ is an external magnetic field, and $x_k$ is the state of the spin at site $k$, which takes values $+1$ for spin up, $-1$ for spin down.
At temperature $T$, the spin configurations at thermal equilibrium obey the Boltzmann distribution, such that
\begin{eqnarray}
\Pr{X_{-N:N}} \propto \mathrm{e}^{-\mathcal{H}_N/k_BT}.
\label{eq:ProbDistFinite}
\end{eqnarray}
where $X_{-N:N} = X_{-N},\ldots X_{N-1}, X_{N}$ and $X_k$ is the random variable that governs $x_k$.

The statistical complexity of such Ising chains has been studied in the thermodynamic limit $N\rightarrow \infty$~\cite{crutchfield1997statistical}.
In this framework we consider a device that scans the chain from left to right.
Suppose at time $t$ the device measures spin $x_t$.
Let $\Lx_t$ govern the statistics of all spins measured so far, to the left of the device, and  $\Rx_t$ all spins to the right of the device yet to be measured.
The probability distribution $\Prline{\Lx_t,\Rx_t}$ then represents the total statistics of what the device observes.
The resulting {\em statistical complexity} then captures the minimum memory requirement of the device which, given access to the state of spins $\lx_t$  left of its current site, can accurately sample from the conditional distribution $\cPrline{\Rx_t}{\Lx_t=\lx_t}$ to simulate spin statistics it has yet to see.
As the spins in the chain are correlated, we expect $\cPrline{\Rx_t}{\Lx_t} \neq \Prline{\Rx_t}$, and thus the memory requirement to be strictly non-zero.
In the thermodynamic limit, $\Prline{\Lx_t,\Rx_t}$ becomes a stationary stochastic process that is invariant with respect to $t$.
Thus we can drop the index $t$ and adopt the standard tools of computational mechanics.

The associated $\epsilon$-machine can be systematically constructed.
Here we summarize prior work~\cite{crutchfield1997statistical}.
A key observation is that $\Prline{\Lx,\Rx}$ is Markovian.
Furthermore, $\cPrline{\Rx}{x_0=+1} \neq \cPrline{\Rx}{x_0=-1}$ for any finite $T$. Thus the Ising system has exactly two causal states, $s_0$ and $s_1$, coinciding respectively with the set of pasts where the device last observed $+1$ or $-1$.
Once initialized in the correct causal state, the resulting $\epsilon$-machine can replicate the correct conditional future statistics according to figure~\ref{emachine}.


\inlineheading{Quantum Statistical Complexity.} A number of approaches to reducing the entropy of classical models using quantum processing have been previously proposed~\cite{gu2012quantum,mahoney2016occam}, but the question of whether such models are optimal has not yet been formally studied. To evaluate the quantum statistical complexity, we first need to define quantum models in general, and then define optimal quantum models.

\begin{definition}[Quantum models] A quantum model is an ordered triple $\mathcal{Q} = (f, \Omega, \mathcal{M})$, where $\Omega$ is a set of quantum states, $f$ is a encoding function that that maps each possible past $\lx$ to a quantum state $\ket{s_{\lx}}\in \Omega$, and $\mathcal{M}$ is a quantum procedure (that is, a completely-positive trace-preserving map) such that application of $\mathcal{M}$ on a physical system $\Xi$ in state $f(\lx)$ should (i) generate an output $x$ with probability $P(X = x|\Lx = \lx)$ and (ii) transition $\Xi$ into state $f (\lx')$, where $\lx'= \lx x = \ldots x_{-1}x_{0} x$ is the resulting past after $x$ is observed.
\end{definition}

Condition (i) ensures that the quantum model generates statistically identical future predictions to its classical $\varepsilon$-machine counterpart. Meanwhile (ii) ensures that the model's internal state updates accordingly at each time-step such that it remains `synchronized,' so that the internal state of machine at any point in time always correctly encodes the past. In this case a series of $L$ repeated application of $\mathcal{M}$ acting on a physical system $\Xi$ initially in state $f(\lx)$, will generate a random variable $X_{1:L}|\lx := X_1 \dots X_L|\lx$ with probability distribution $P(X_{1:L} | \Lx = \lx)$, for any desired $L \in \mathbb{Z}^+$. In the limit $L\!\rightarrow\!\infty$ the quantum model will therefore replicate the process's future behaviour by generating a random variable $\future{X}$ with probability distribution $P(\Rx | \Lx = \lx)$, whenever it is supplied with a system $\Xi$ in state $f(\lx)$. This definition assumes encoded states to be pure to maintain unifilarity, in line with classical models.
However, the results we shall present here continue to hold even if this assumption is dropped (see Lemma \ref{lem:PureCloseStates} in appendix).

The entropy $\Ent{\rho}$ of the stationary mixture of encoded states $\rho$ represents the amount of past information the model must store to correctly simulate the future.
This motivates
\begin{equation}
C(\mathcal{Q}) =  \Ent{\rho} = -\text{Tr}(\rho \log{\rho}),
\end{equation}
as the \emph{complexity} of a quantum model,
 where $\rho=\sum_i p_i \ket{s_i}\bra{s_i}$ when $p_i = \Prline{f(\lx)\!=\!\ket{s_i}}$ is the probability of the model being in the quantum state $\ket{s_i}$.
An optimal quantum model is then the quantum model with the least complexity:

\begin{definition}[Optimality] A quantum model $\mathcal{Q}$ of a process $\Prline{\Lx,\Rx }$ is optimal for $\Prline{\Lx,\Rx }$ if and only if for any other quantum model $\mathcal{Q}'$ of $\Prline{\Lx,\Rx }$, $C(\mathcal{Q}) \leq C(\mathcal{Q}')$.
\end{definition}

Unlike the classical $\epsilon$-machine, there may be more than one optimal quantum model. At the very least, for any optimal quantum model $\mathcal{Q}_0$ with a set of states $\Omega_0$, one can form another quantum model $\mathcal{Q}_0'$ with states $\Omega_0'$ trivially related to those in $\Omega_0$ by a unitary transformation.
Since the von Neumann entropy is invariant under unitary transformations, $\mathcal{Q}_0'$ will also be an optimal model.

\begin{definition}[Quantum Causal States]
If $\mathcal{Q}_0=\left( f_0,\Omega_0,\mathcal{M}_0 \right)$ is an optimal quantum model for a process $\Prline{\Lx,\Rx}$, then $\Omega_0$ are a set of {\em quantum causal states} for $\Prline{\Lx,\Rx}$.
\end{definition}

This furnishes the necessary background for a definition of quantum statistical complexity.

\begin{definition}[Quantum Statistical Complexity]
Consider a stochastic process $\Prline{\Lx,\Rx }$ with an optimal quantum model $\mathcal{Q}_0$.
Let $\Ent{\rho}$ be the amount of information stored by $\mathcal{Q}_0$.
The quantum statistical complexity of $\Prline{\Lx,\Rx }$ is defined as $C_q = \Ent{\rho}$.
\end{definition}

Thus the quantum statistical complexity represents the minimal past information an optimal quantum model must record in order to accurately sample from \mbox{$\cPrline{\Rx}{\Lx\!=\!\lx}$}. Evaluating the quantum statistical complexity is generally non-trivial. While general methods of constructing better-than-classical quantum models are known~\cite{gu2012quantum,mahoney2016occam,riechers2015closed}, it remains an open question if and when such models are optimal. Thus to date, $C_q$ has never been computed. We do not know in general exactly how much simpler a process can be when modelled quantum mechanically, nor if the quantum statistical complexity exhibits qualitatively different behaviour.
\begin{figure}[t!]
\centering
\includegraphics[width = 0.5\textwidth]{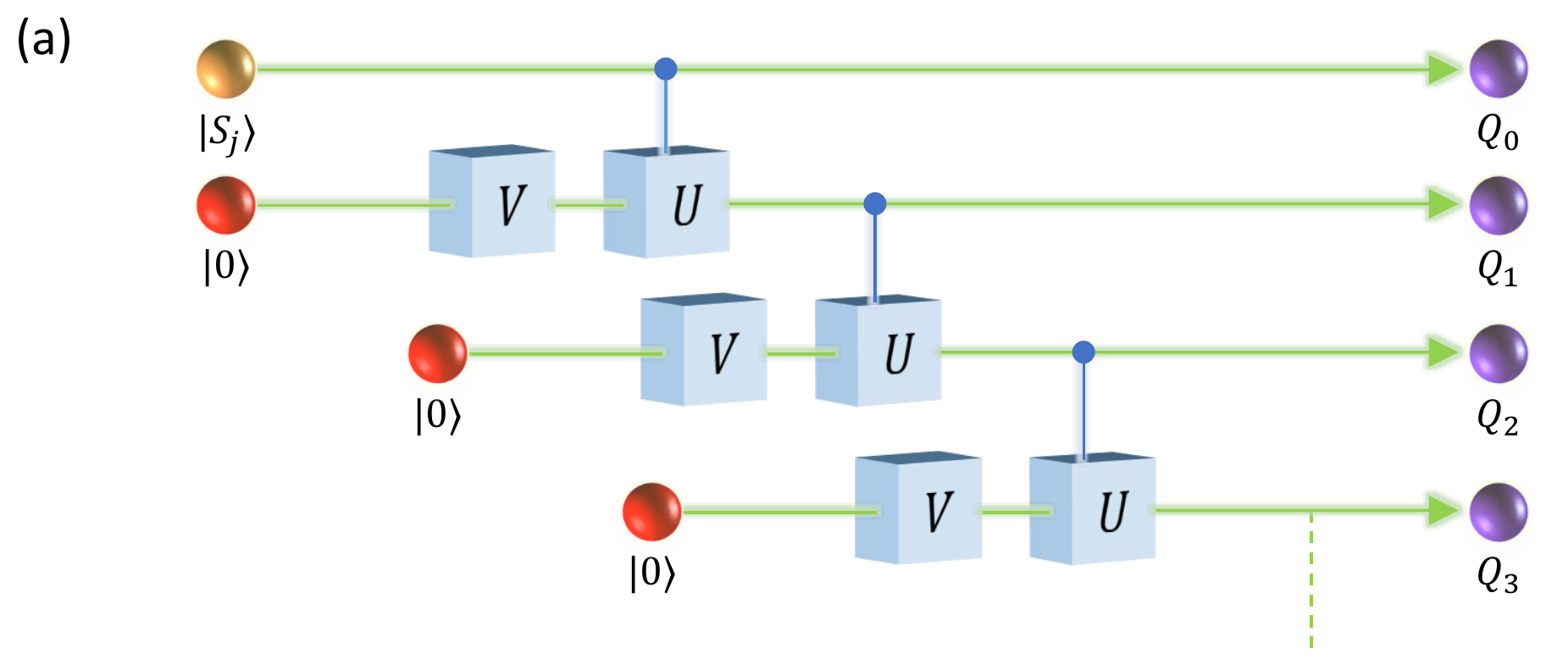}
\includegraphics[width = 0.5\textwidth]{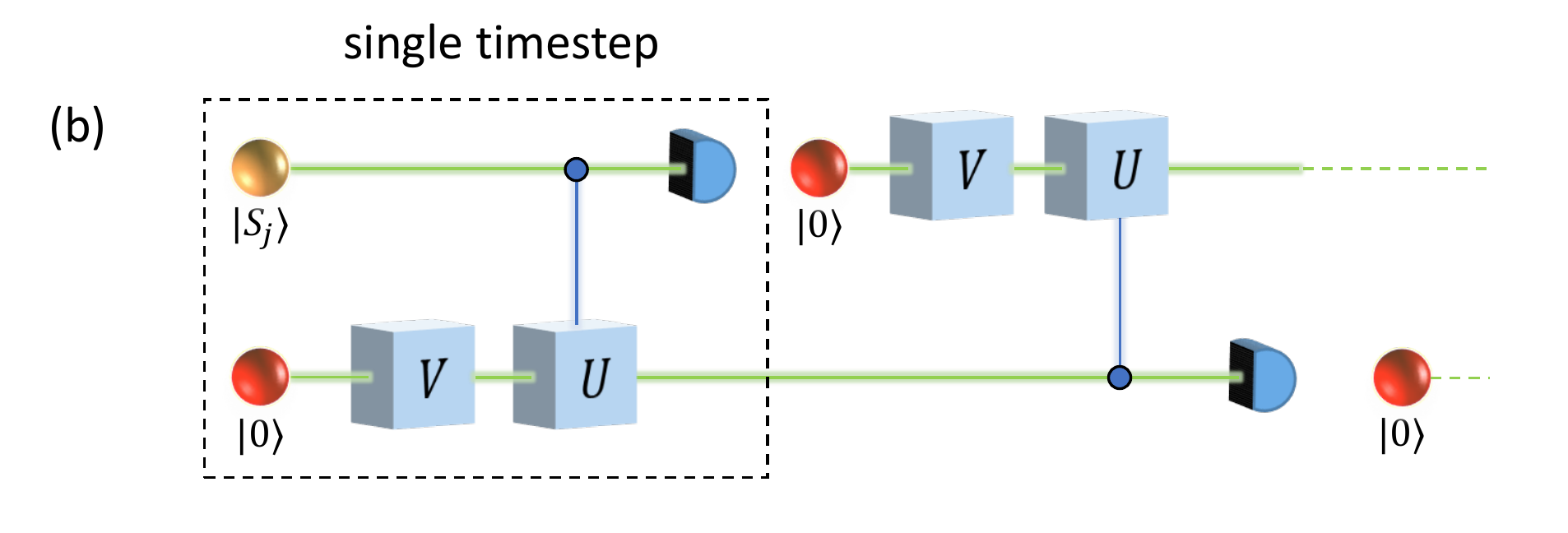}
\\
\caption{(a) Quantum circuit of how the quantum $\epsilon$-machine can sample from the desired probability distribution $\cPrline{\Rx}{\Lx = \rx}$ when given a quantum causal state $\ket{s_j} = \epsilon_q(\lx)$. We introduce two unitaries: $V$ that maps $\ket{0}$ to $\ket{s_0}$, and $U$ such that $U \ket{s_0} = \ket{s_1}$. The machine is first supplied with an ancillary qubit that is systematically initialized in state $\ket{s_0}$ by application of $V$. Subsequently, it applies $U$ on the ancillary qubit, controlled on the memory (system in state $\ket{s_i}$), where the control unitary transformation maps $\ket{k}\ket{\phi} \rightarrow \ket{k} U^k \ket{\phi}$ and $k\in\{0,1\}$ indexes the computational basis. The original causal state is then emitted from the machine (denoted by $Q_0$). Repetition of this procedure yields a large entangled chain of qubits $Q_1,Q_2,\ldots$ that contains a superposition of all possible futures. Measurement of $Q_k$ with respect to observable $\mathcal{Z}=\ket{0}\bra{0} - \ket{1}\bra{1}$ then yields the correct statistics for $x_k$. (b) Alternatively, we can view the above circuit as an iterative procedure. We can measure each as soon as it is emitted, which will collapse the corresponding ancillary qubit to the desired quantum causal state for the subsequent timestep.
 }
\label{qcircuit}
\end{figure}

Here we aim to construct a provably optimal quantum model for the Ising spin chain. To do this, we first introduce two useful lemmas.

\begin{lemma}[Causal state correspondence]
\label{lem:OneToOne}
For any stochastic process $\Prline{\Lx,\Rx }$, there exists an optimal quantum model $\mathcal{Q}_0 = \{ f_0, \Omega_0, \mathcal{M}_0\}$ of $\Prline{\Lx,\Rx }$ such that for any two pasts $\past{x}$, $\past{x}'$,  $f_0\!\left(\past{x}\right) = f_0\!\left(\past{x}'\right)$ if and only if $\epsilon\!\left(\past{x}\right) = \epsilon\!\left(\past{x}'\right)$.
\end{lemma}
That is, there exists an optimal quantum model that has a one-to-one correspondence between the causal states and quantum causal states.
This lemma implies there is never any benefit in differentiating two different pasts with coinciding conditional futures.
The formal proof is based on the concavity of the von Neumann entropy, and is supplied in the technical appendix.

The second lemma places a constraint on how similar two quantum causal states can be, if we require the model to still be capable of generating correct future output statistics.

\begin{lemma}[{Maximum fidelity constraint}]
\label{lem:Minimum} Consider a stochastic process $ \Prline{\Lx,\Rx }$ with a valid
 a quantum model $\mathcal{Q} = \{f, \Omega, \mathcal{M}\}$,
 where $\Omega = \{\ket{s_i}\}_i$ consists of quantum states in one-to-one correspondence with classical causal states $\{s_i\}_i$. Let $\sigma_i = \ket{s_i}\bra{s_i}$. Then the following statement is true: $F\!\left(\sigma_i, \sigma_j\right) \leq F^{ij}_{\rm max}$ for all $i$ and $j$,
where $F$ is the Fidelity and $F^{ij}_{\rm max} = F\!\left(\cPrline{\future{X}}{s_i}, \cPrline{\future{X}}{s_j} \right)$\footnote{We remind readers that the fidelity between two classical probability distributions is a special case of the fidelity between two quantum states, hence we use the same symbol for both cases.}.
\end{lemma}
The formal proof follows from the monotonicity of the fidelities under valid quantum operations, and is supplied in the technical appendix.

\inlineheading{An optimal quantum model of the spin chain.}
We now apply these general results to the 1D Ising spin chain.
To discuss the quantum statistical complexity, we must identify an optimal quantum model of the process.
We shall proceed by presenting a model $\mathcal{Q}_0 = (\epsilon_q, \Omega_0, \mathcal{M}_0)$, based on the construction in~\cite{gu2012quantum}.
We will later prove that this model is optimal for the Ising spin chain, and hence refer to $\mathcal{Q}_0$ as the {\em quantum $\epsilon$-machine} for this process.
This model uses two quantum states $\Omega_0 = \{\ket{s_0},\ket{s_1}$\} where
\begin{eqnarray}
\ket{s_0}&=&\sqrt{T_{00}}\ket{0} + \sqrt{T_{01}}\ket{1}, \nonumber \\
\ket{s_{1}}&=&\sqrt{T_{10}}\ket{0} + \sqrt{T_{11}}\ket{1} \label{qcausal}
\end{eqnarray}
and $T_{ij}$ is the transition probability for the classical $\epsilon$-machine of the process in state $s_i$ to output $(-1)^j$ and transition to $s_j$.
The associated encoding map is $\epsilon_q: \lx \rightarrow \ket{\epsilon(\lx)}$, where $\epsilon$ is the corresponding map from $\lx$ to classical causal states.
As shown explicitly in figure \ref{qcircuit}, there exists a quantum procedure $\mathcal{M}_0$ that allows this model to systematically sample from $\cPrline{X_1\ldots X_L}{\Lx\!=\!\lx}$ for any $L$, by repeated applications of $\mathcal{M}_0$ acting on a physical system $\Xi$ initialized in state $\epsilon_q(\lx)$.
Thus the resulting triple $\mathcal{Q}_0 = (\epsilon_q, \Omega_0, \mathcal{M}_0)$ is indeed an accurate quantum model for the Ising spin chain.

This model has an internal entropy
\begin{align}
\Ent{\rho}&= -\Tr\left( \rho \log{\rho} \right), \label{epsilonentropy}
\end{align}
where
\begin{eqnarray}
\rho=p_0\ket{s_0}\bra{s_0} + p_1\ket{s_1}\bra{s_1}
\end{eqnarray}
such that $p_0$ $( p_1 )$ is the probability of any spin being up (down).
Since $\ket{s_0}$ and $\ket{s_1}$ are generally not mutually orthogonal, we see immediately that this machine stores less information than any possible classical model.
That is $C_q \leq \Ent{\rho} < C_\mu$.

To establish that $\Ent{\rho}$ in indeed the quantum statistically complexity (i.e.\ that $\Ent{\rho}=C_q$), we need to establish that $\mathcal{Q}_0$ is an optimal quantum model for the Ising system.

\begin{result}The quantum $\epsilon$-machine
$\mathcal{Q}_0 = (\epsilon_q, \Omega_0, \mathcal{M}_0)$
is an optimal quantum model for the 1D Ising spin chain, and hence its internal entropy $\Ent{\rho}$ corresponds to the quantum statistical complexity $C_q$ of this system.
\end{result}
\vspace{-0.5em} The proof is rather involved, and we present the full details in the technical appendix.
This proof relies on the Markovian nature of the 1D Ising system, and that it only has two casual states.
The basic approach is to first note that it is sufficient to examine quantum models with two states (Lemma~\ref{lem:OneToOne}),
 and then directly evaluate the fidelity $F\small[\cPrline{\future{X}}{s_0}, \cPrline{\future{X}}{s_1} \small]$ for the spin chain and show that it coincides with $F(\ket{s_0}\bra{s_0}, \ket{s_1}\bra{s_1})$.
Lemma \ref{lem:Minimum} is then invoked together with the monotonic relationship between the fidelity of two\footnote{
This monotonic relationship does not hold for mixtures of more than two states~\cite{jozsa2000}. This presents a significant challenge for generalizations of this proof to processes with more than two causal states.
} quantum states and the entropy of their statistical mixture to establish optimality of $\mathcal{Q}_0$.

This result formally establishes equation~\eqref{epsilonentropy} as an analytical expression for $C_q$, allowing us to compare its qualitative behaviour with $C_\mu$.

\begin{figure}[tbh]
	\centering
		\includegraphics[width = 0.5\textwidth]{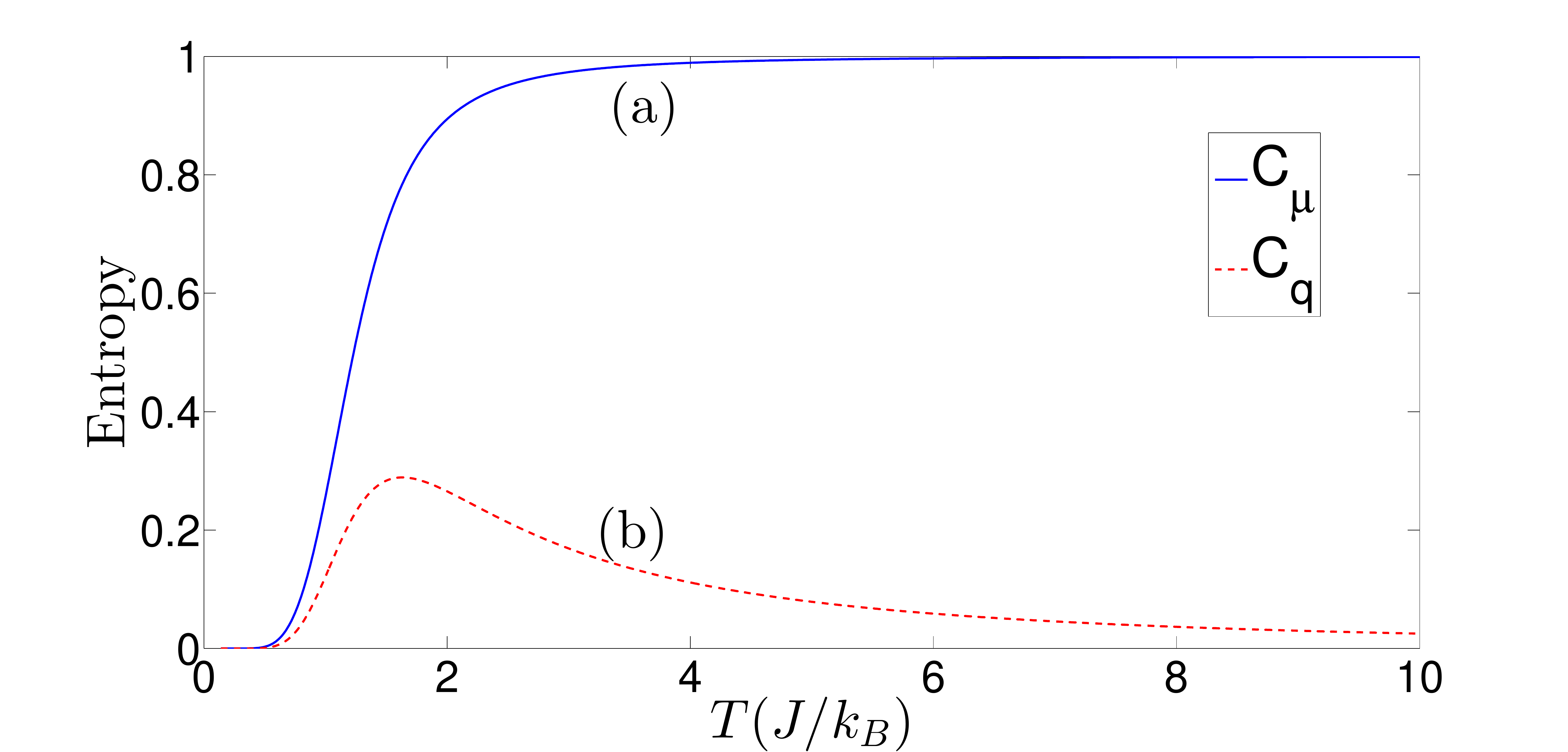}
		\caption{Plot of classical and quantum statistical complexity for various temperature of the Ising spin chain, with coupling parameter $J=1$ and the magnetic field strength $B=0.3$. Whereas the classical statistical complexity (a) monotonically increases with temperature, its quantum counterpart (b) attains maximal value for finite $T$.}
		\label{plot1}
\end{figure}

\textbf{Qualitative Divergences.} The quantum statistical complexity and its classical counterpart exhibit very different qualitative behaviour (figure~\ref{plot1}). The classical statistical complexity $C_\mu$ increases monotonically with temperature for all finite temperatures $T$. This reflects the fact that at hotter temperatures, all spin states become more equally likely. Thus the two classical causal states become progressively more equiprobable, resulting in a higher associated memory cost. Note that at infinite temperature, $C_\mu$ drops discontinuously to zero. This can be understood through the observation that at this limit, the spins observed on the left become completely uncorrelated with those on the right, and thus no memory is necessary to simulate the spin chain.

In contrast, the quantum statistical complexity $C_q$ peaks at some finite temperature $T_{\text{max}}$.
Progressive \emph{increases} in temperature past this point result in a decrease in the amount of quantum memory required -- $C_q$ decays smoothly to $0$ as $T \rightarrow \infty$. We conclude that for $T > T_{\text{max}}$, the quantum and classical analogues of statistical complexity diverge in qualitative behaviour.
The Ising system takes progressively more memory to simulate classically with increasing temperature, this relation is reversed in quantum models for high temperatures.
Indeed, as $T \rightarrow \infty$, the efficiency ratio $C_\mu/C_q$ becomes unbounded. 

At this stage, we note an important caveat. The statistical complexity is an entropic quantity, and thus takes operational meaning in the independent identically-distributed limit.
Specifically, suppose a processes has respective classical and quantum statistical complexity of $C_\mu$ and $C_q$. This implies that if we wish to replicate the futures of an ensemble of $N$ such processes in the limit of large $N$, then we would either need $N C_\mu$ bits or $N C_q$ qubits. However, if we are simulating only a single instance of a stochastic process, $C_q < C_\mu$ does not necessarily imply that the process can be simulated with less qubits than bits. Indeed, modelling a single instance of the spin chain would require either $1$ bit or $1$ qubit. It would therefore be interesting to see if these qualitative divergences can exist using single shot measures, such as a max entropy of the input (known in complexity literature as the topological state complexity~\cite{crutchfield1989inferring}).

\inlineheading{Discussion.}  The statistical complexity -- the amount of historical information we must retain about a system's past in order to simulate its future behaviour -- is a popular operational quantifier of structure and complexity. Here we study how this measure can change when quantum processing is allowed. Using a 1D Ising system, we showed that the classical and quantum statistical complexity can diverge drastically in qualitative behaviour. Whereas classical models identify spin chains at very high temperatures as the most memory intensive to simulate and thus the most complex, quantum models identify these as being very simple. Our techniques also establish two necessary properties that any optimal quantum model must satisfy, which contribute towards computing quantum statistical complexity in more general settings.

More foundationally, these results suggest that complexity is an observer-dependent concept. Different observers may construe a system to be highly complex or extremely simple depending on whether they could reason quantum mechanically. This motivates a number of questions: how pervasive are such divergences and how do they impact our current understanding of what is complex? Certainly, the Ising model is used to model diverse phenomena in the physical and biological sciences ~\cite{schneidman2006weak,hill1959generalization,simon1973varphi4,torquato2011toward}. Meanwhile generalisations of statistical complexity have been used to study the dynamics of self-organization~\cite{shalizi2004quantifying} and the structure of input-output processes~\cite{barnett2015computational}, and it would be interesting to see how such results may change when viewed through the lens of quantum theory.

\inlineheading{Note added.} During the process of referee review, further work has highlighted the qualitative difference between $C_q$ and $C_\mu$ leads to ambiguity about the simplest parameter regime for a particular system, and discussed generality of this phenomenon~\cite{Agha16b}. The unbounded efficiency ratio between $C_q$ and $C_\mu$ has also since been shown in the context of progressively longer $N$-nearest neighbour Dyson-Ising systems~\cite{Agha16a} (in the limit of $C_q\to0$, as in this article), as well as in simulating discretizations of continuous systems~\cite{Garner16a} and stochastic processes in continuous time~\cite{elliot17}  (where $C_q$ is bounded, but $C_\mu$ grows unboundedly). Meanwhile, the differences in $C_\mu$ and $C_\mu$ have since being extended to cover input-output processes that react differently to different inputs~\cite{Thompson17}.

\acknowledgements
We thank Felix Binder, Michele Dall'Arno, \mbox{Alex Monras}, Blake Pollard, \mbox{Karoline Wiesner}, and Benjamin Yadin for enlightening discussion.
We are grateful for funding from  the National Research Foundation of Singapore and in particular NRF Award No. NRF--NRFF2016--02 and NRF CRP Award No. NRF-CRP14-2014- 02;
 the John Templeton Foundation Grant 53914 {\em ``Occam's Quantum Mechanical Razor: Can Quantum theory admit the Simplest Understanding of Reality?''};
 the Foundational Questions Institute,
 the Ministry of Education in Singapore, the Academic Research Fund Tier 3 MOE2012-T3-1-009;
 the National Natural Science Foundation of China Grants 11450110058, 61033001 and 61361136003;
 the 1000 Talents Program of China;
 and the Oxford Martin School.

\newpage
\section*{TECHNICAL APPENDIX}
\inlineheading{The $\epsilon$-machine for Ising spin chains.}
Let $\mathcal{P} =\Pr{\Lx,\Rx}$ denote the probability distribution of an Ising spin chain.
Crutchfield and Feldman~\cite{crutchfield1997statistical} show that
\begin{eqnarray}
\cPr{\Rx}{\Lx = \lx}&=&\cPr{\Rx}{X_{0} = x_{0}},
\end{eqnarray}
that is, the probability of observing the right half spin configurations given the knowledge of the left half spin configuration, requires only the knowledge of $x_{0}$\cite{feldman1998computational}. This implies that any two pasts $\lx = \ldots,x_{-1},x_{0}$, $\lx' = \ldots,x'_{-1},x'_{0}$ belong to the same causal state provided $x_0 = x_0'$.
Thus, the $\epsilon$-machine for this process has two causal states, $s_j = \{\lx: x_{0} = -1^j\}$, with $j = 0, 1$.
The probability of finding the system in either causal state is thus given by the probability of finding any spin in the system to be up or down, denoted by $p_0$ and $p_1$ respectively.

As the causal states of the system are in one to one correspondence with $x_{0}$, $T^{(r)}_{ij}$ is non-zero only when $r = (-1)^j$.
Thus, the notation for the $\epsilon$-machine's dynamics can be simplified by writing $T^{(+1)}_{i0} = T_{i0}$ and $T^{(-1)}_{i1} = T_{i1}$.
Furthermore the exact transition probabilities between causal states
\begin{eqnarray}
T_{ij} = \cPr{X_{k+1} = (-1)^j}{X_{k} = (-1)^i}
\end{eqnarray}
align with the probabilities governing the spin at location $k+1$ given the state of the spin at location $k$.
The exact values of these transition probabilities vary as a function of temperature, and are well known~\cite{yeomans1992statistical}.


\begin{replemma}{lem:OneToOne}[{Causal state correspondence}]
For any stochastic process $\Prline{\Lx,\Rx }$, there exists an optimal quantum model $\mathcal{Q}_0 = \{ f_0, \Omega_0, \mathcal{M}_0\}$ of $\Prline{\Lx,\Rx }$ such that for any two pasts $\past{x}$, $\past{x}'$,  $f_0\!\left(\past{x}\right) = f_0\!\left(\past{x}'\right)$ if and only if $\epsilon\!\left(\past{x}\right) = \epsilon\!\left(\past{x}'\right)$.
\begin{proof}

We first prove the ``only if'' by contradiction.
Consider two pasts $\past{x}$ and $\past{x}'$ where $\epsilon\!\left(\lx\right) \neq \epsilon\!\left(\lx'\right)$.
This implies $\cPrline{\Rx}{\lx} \neq \cPrline{\Rx}{\lx'}$.
For some quantum model $\mathcal{Q} = \{ f, \Omega, \mathcal{M}\}$,
 let $\ket{\phi_{\past{x}}} = f\!\left(\past{x}\right)$ and $\ket{\phi_{\past{x}'}} = f\!\left(\past{x}'\right)$.
Set density matrices $\sigma_{\past{x}} = \ketbra{\phi_{\past{x}}}{\phi_{\past{x}}}$ and $\sigma_{\past{x}'} = \ketbra{\phi_{\past{x}'}}{\phi_{\past{x}'}}$.
Now if $\sigma_{\past{x}} = \sigma_{\past{x}'}$, then there cannot exist a systematic procedure $\mathcal{M}$ that can output different statistics on input of either $\ket{\phi_{\past{x}}}$ or $\ket{\phi_{\past{x}'}}$.
Thus no quantum model can generate the correct conditional statistics for both $\lx$ and $\lx'$.
Therefore for any quantum model that simulates $\mathcal{P}$, $\sigma_{\past{x}} \neq \sigma_{\past{x}'}$ whenever $\epsilon\!\left(\lx\right) \neq \epsilon\!\left(\lx'\right)$.

The ``if'' direction will be demonstrated using concavity of entropy. We show that for an arbitrary quantum model $\mathcal{Q}'$, one can always find a quantum model $\mathcal{Q}$ with $C(\mathcal{Q}) \leq C(\mathcal{Q'})$ which satisfies $\sigma_{\past{x}} = \sigma_{\past{x}'}$ if $\epsilon\!\left(\past{x}\right) = \epsilon\!\left(\past{x}'\right)$.

Consider the set of pasts in the same classical causal state $s_i$.
Let $P_i = \Pr{\lx \in s_i} > 0$ be the total probability in the stationary state that the past is in causal state $s_i$.
Let $\rho$ be the stationary state of quantum model $\mathcal{Q}'$,
 which we divide into contributions from pasts in $s_i$ and pasts not in $s_i$: $\rho = \sum_{\past{x} \notin s_i} \Prline{\past{x}} \sigma_{\past{x}} + \sum_{\past{x} \in s_i} \Prline{\past{x}} \sigma_{\past{x}}$, where $\Prline{\past{x}} = \Prline{\Lx = \lx}$.
By writing
 $\rho_{\lnot i} = \frac{1}{1 - P_i} \sum_{\past{x} \notin s_i} p\!\left(\past{x}\right) \sigma_{\past{x}}$,
and $q\!\left(\past{x}\right) = p\!\left(\past{x}\right) / P_i$,
 we write
 $\rho = \left(1\!-\!P_i\right) \rho_{\lnot i} + P_i \sum_{\past{x} \in s_i} q\!\left(\past{x}\right) \sigma_{\past{x}}$,
  and can then factor out the summation such that
 $\rho = \sum_{\past{x} \in s_i} q\!\left(\past{x}\right) \left[  \left(1-P_i\right) \rho_{\lnot i} + P_i \sigma_{\past{x}} \right]$.

From the concavity of entropy,
\begin{align}
\Ent{\rho} \geq & \sum_{\past{x} \in s_i} q\!\left(\past{x}\right) \Ent{ \left(1-P_i\right) \rho_{\lnot i} + P_i \sigma_{\past{x}} } \nonumber \\
\geq & \min_{\past{x} \in s_i} \{ \Ent{ \left(1-P_i\right) \rho_{\lnot i} + P_i \sigma_{\past{x}} } \}.
\end{align}
Let $\past{x}_{m} \in s_i$ be the past that minimizes this bound, with associated density operator $\tilde{\sigma} = \sigma_{\past{x}_{m}}$.
For every quantum model where $\past{x}\in s_i$ are assigned to arbitrary $\sigma_{\past{x}}$, there is also a valid quantum model that assigns the state $\tilde{\sigma}$ for all $\past{x} \in s_i$,
 since for all pasts in $s_i$, repeatedly applying the measurement process $\mathcal{M}$ to $\tilde{\sigma}$ will give the desired future statistics.
The stationary state of this model is $\tilde{\rho} = \left(1-P_i\right) \rho_{\lnot i} + P_i \tilde{\sigma}$, with an entropy of $\Ent{\tilde{\rho}} = \Ent{ \left(1-P_i\right) \rho_{\lnot i} + P_i \tilde{\sigma} } \leq \Ent{\rho}$.
It follows by considering every causal state $s_i$ in turn, one can construct a quantum model with lower or equal entropy that maps all $\past{x}\in s_i$ to the same quantum state $\sigma_i$.

Hence, for any arbitrary quantum model, there is always another valid quantum model with the same or lower entropy that satisfies $\sigma_{\past{x}} = \sigma_{\past{x}'}$ if and only if $\epsilon\!\left(\past{x}\right) = \epsilon\!\left(\past{x}'\right)$.
As such, we can restrict our search for an optimal model to those that satisfy the criterion of this lemma.

Finally, we must establish if there is a valid quantum model at all.
This is guaranteed by the existence of a classical $\epsilon$-machine, which is a special case of quantum models (with the causal states encoded onto orthogonal quantum states).

\end{proof}
\end{replemma}

\begin{replemma}{lem:Minimum}[Maximum fidelity constraint]
Consider a stochastic process $ \Prline{\Lx,\Rx }$ with a valid
 a quantum model $\mathcal{Q} = \{f, \Omega, \mathcal{M}\}$,
 where $\Omega = \{\ket{s_i}\}_i$ consists of quantum states in one-to-one correspondence with classical causal states $\{s_i\}_i$. Let $\sigma_i = \ket{s_i}\bra{s_i}$. Then the following statement is true: $F\!\left(\sigma_i, \sigma_j\right) \leq F^{ij}_{\rm max}$ for all $i$ and $j$,
where $F$ is the Fidelity and $F^{ij}_{\rm max} = F\!\left(\cPrline{\future{X}}{s_i}, \cPrline{\future{X}}{s_j} \right)$.
\begin{proof}
The proof of follows from the monotonicity of $F$ under trace preserving quantum operation~(see, e.g.\ chapter 9 of \cite{nielsen2010quantum}). A measurement $\mathcal{M}$ that extracts statistics must induce a quantum operation $\sigma_i \mapsto \mathcal{R}(\sigma_i)$. Let $\ket{\psi}$ and $\ket{\phi}$ be purifications of $\sigma_i$ and $\sigma_j$ into a joint system $AB$ such that $\sigma_i = \text{tr}_B(\ket{\psi}\bra{\psi})$, $\sigma_j = \text{tr}_B(\ket{\phi}\bra{\phi})$, and $F(\sigma_i,\sigma_j)=|\braket{\psi}{\phi}|$. Next, we introduce an environment $E$ for the quantum operation $\mathcal{R}$, such that $\mathcal{R}(\sigma_i)=\text{tr}_E\left( U\left( \sigma_i\otimes\ket{0}\bra{0} \right)U^\dagger \right)$. We then note that $\mathcal{R}(\sigma_i)=\text{tr}_{BE}\left( U\left( \ket{\psi}\bra{\psi}\otimes \ket{0}\bra{0} \right)U^\dagger \right)$. This also applies to $\mathcal{R}(\sigma_j)$. By Uhlmann's theorem~\cite{nielsen2010quantum},
\begin{eqnarray}
F\left( \mathcal{R}(\sigma_i), \mathcal{R}(\sigma_j) \right) &\geq& |\bra{\psi}\bra{0}U^\dagger U \ket{\phi}\ket{0} | \\
 &=& |\braket{\psi}{\phi}| \\
 &=& F(\sigma_i,\sigma_j).
\end{eqnarray}
Hence, there is a maximum fidelity $F_\text{max}$ between the quantum states $\sigma_i$ and $\sigma_j$ set by $F_\text{max}=F\!\left(\cPrline{\future{X}}{s_i}, \cPrline{\future{X}}{s_j} \right)$. If this bound is violated, $\mathcal{Q}$ is not a valid quantum model.
\end{proof}
\end{replemma}

\inlineheading{Optimality of the Quantum $\epsilon$-Machine.}
We prove the following theorem:
\begin{theorem*} \label{thm:optimal}
Let  $\mathcal{P} = \Prline{\Rx,\Lx}$ denote the stochastic process that governs that Ising system,
 and $s_0$ and $s_1$ be the causal states of this process, with transition probabilities $T_{ij}$ in the corresponding $\epsilon$-machine.
Then there is an optimal quantum model $\mathcal{Q}_{0} = \{\epsilon_q, \Omega_0, \mathcal{M}_0\}$ where $\Omega_0$ consists
of the pure quantum states $\ket{s_0}$ and $\ket{s_0}$, given by
\begin{align}
\ket{s_0}&=\sqrt{T_{00}}\ket{0} + \sqrt{T_{01}}\ket{1}, \nonumber \\
\ket{s_1}&=\sqrt{T_{10}}\ket{0} + \sqrt{T_{11}}\ket{1}. \label{qcausal1}
\end{align}

\begin{proof}
From Lemma \ref{lem:OneToOne}, we know that there is an optimal quantum model with two quantum causal states.
Let them be $\ket{\psi_0}$ and $\ket{\psi_1}$.
The complexity of this model monotonically decreases with increasing fidelity $F(\ketbra{\psi_0}{\psi_0}, \ketbra{\psi_1}{\psi_1})$ between the two causal states. From Lemma \ref{lem:Minimum}, we know that $F(\ketbra{\psi_0}{\psi_0}, \ketbra{\psi_1}{\psi_1}) \leq F_{\rm max} = F\!\left(\cPrline{\future{X}}{s_0}, \cPrline{\future{X}}{s_1} \right)$ for any valid quantum model.
Therefore, if there is valid quantum model such that the fidelity $F(\ketbra{\psi_0}{\psi_0}, \ketbra{\psi_1}{\psi_1}) = F_{\rm max}$, then this model is optimal.

The fidelity between the two distributions $\cPrline{\Rx}{s_0}$ and $\cPrline{\Rx}{s_1}$ is given by
\begin{align}
F(\cPr{\rx}{s_0},\cPr{\rx}{s_1}) & \nonumber\\
&\hspace{-7em} = \sum_{\rx_i}{\sqrt{\cPr{\Rx=\rx_i}{s_0}\cPr{\Rx=\rx_i}{s_1}}} \nonumber \\
& \hspace{-7em} = \sum_{x_0,x_1,\ldots}\sqrt{\cPr{x_0,x_1,\ldots}{s_0}} \sqrt{\cPr{x_0,x_1,\ldots}{s_1}},
\end{align}
where $x_k$ represents the processes' output at time $t=k$.
The Ising system is a Markovian process where $T^{(r)}_{ij} = T^{(r)}_{ij}\delta_{j\frac{1-r}{2}}$, allowing us to express the transition probabilities as $T_{ij}$. Under these circumstances we can invoke the chain rule of probability and re-express:
\begin{flalign*}
F(\cPr{\rx}{s_0}, \cPr{\rx}{s_1}) &&
\end{flalign*}
\vspace{-2em}
\begin{align}
 = &  \sum_{x_1,x_2,\dots} \sqrt{\cPr{x_1}{s_0} \cPr{x_2}{x_1}\ldots} \nonumber \\
& \hspace{3.5em} \times \sqrt{\cPr{x_1}{s_1}\cPr{x_2}{x_1}\ldots} \nonumber \\
= &  \sum_{x_1} \sqrt{\cPr{x_1}{s_0} \cPr{x_1}{s_1}} \sum_{x_2,x_3,\ldots} \hspace{-0.5em} \cPr{x_2, x_3 \dots}{x_1} \nonumber \\
=& \sum_{x_1} \sqrt{\cPr{x_1}{s_0}\cPr{x_1}{s_1}} \nonumber \\
= & ~\sqrt{T_{00}T_{10}} + \sqrt{T_{01}T_{11}}. \label{innerbound}
\end{align}

It can be easily verified that for the quantum states $\ket{s_0}$ and $\ket{s_1}$ presented in equation~\eqref{qcausal1}, $F(\ketbra{s_0}{s_0}, \ketbra{s_1}{s_1}) = \sqrt{T_{00}T_{10}} + \sqrt{T_{01}T_{11}}$, and hence these states saturate the maximal possible fidelity for $\mathcal{P}$.

As such, the only remaining step is to demonstrate the existence of an $\mathcal{M}$ that extracts the necessary statistics.
This is presented constructively by the quantum circuit in figure~\ref{qcircuit} of the text.
Hence, $\mathcal{Q}_0$ is a valid quantum model, and moreover, since it consists of two pure states that saturate the maximum fidelity, it is an optimal quantum model.
\end{proof}
\end{theorem*}

Since we have found an optimal quantum model $\mathcal{Q}_{0}$ for the Ising spin chain, this demonstrates that the spin chain's quantum statistical complexity $C_q = C(\mathcal{Q}_{0})$.
We remark that there is in fact a family of optimal models.
Fixing the first quantum causal state $\ket{\sigma_0}$ as $\ket{0}$ without loss of generality (by symmetry),
 we can write the second quantum causal state
 $\ket{\sigma_1} = \cos\left(\frac{\theta}{2}\right) \ket{0} + e^{i \phi} \sin\left(\frac{\theta}{2}\right) \ket{1}$,
 where  $\theta \in (0,\pi]$ is determined by the maximum fidelity,
 but we are free to choose any $\phi \in [0, 2\pi)$ and any $\ket{1}$ such that $\braket{0}{1} = 0$.
All such assignments of $\ket{\sigma_0}$ and $\ket{\sigma_1}$ are all related by unitary transformations,
 so if there is a measurement $\mathcal{M}$ that extracts the statistics from one such model, a measurement can be found for all such models.
Moreover, since the von Neumann entropy is basis-independent, all such choices of model have the same complexity.
It thus follows that every assignment of pure states separated by the minimum distance are optimal.

We also establish that the Main Theorem continues to hold when we consider encoding $\lx$ into mixed quantum states, by the following lemma:
\begin{lemma}
\label{lem:PureCloseStates}
Consider a more general quantum model $\tilde{\mathcal{Q}} = \{\tilde{f}, \tilde{\Omega}, \tilde{\mathcal{M}}\}$, where the encoding map $\tilde{f}$ takes each $\lx$ to some possibly mixed state $\sigma_{\lx} \in \tilde{\Omega}$. (Clearly, the set of all $\mathcal{Q}$ forms a subset of the set of $\tilde{\mathcal{Q}}$.)
Let $P(\past{X},\future{X})$ denote a Markovian stochastic process with two causal states $s_0$ and $s_1$, and two possible outputs. If there exists a quantum model $\mathcal{Q}_0 = \{f_0, \Omega_0, \mathcal{M}_0\}$ where $\Omega_0$ consists of two pure quantum states $\sigma_0$ and $\sigma_1$ (in one-to-one correspondence with classical causal states), saturating the maximum fidelity  $F_{\rm max} =  F\!\left(\cPrline{\future{X}}{s_0}, \cPrline{\future{X}}{s_1} \right)$,
then this $\mathcal{Q}_0$ is optimal.

\begin{proof}
It has already been established by the Main Result that such a $\mathcal{Q}_0$ is optimal amongst pure state quantum models.
We shall show that there are no alternative models with a lower entropy than $\mathcal{Q}_0$, even when we allow for encoding onto mixed quantum states.

First, note that the argument of Lemma~\ref{lem:OneToOne} trivially generalizes to mixed states, we may restrict our search within models with two internal states.

Let $\mathcal{Q}'$ be a quantum model with $\Omega' = \{\tilde{\sigma_0},\tilde{\sigma_1}\}$,  such that at least one state (say, $\tilde{\sigma}_1$) is mixed.
As $\mathcal{Q}'$ is a valid quantum model for $P(\past{X},\future{X})$, there is a measurement process that yields outcome $x_1 \in \{-1,1\}$ with probability $P(X_1 = x_1 | S_{0} = s_i)$ when acting on $\tilde{\sigma_i}$.
Let this measurement have elements $\{M, \mathbb{I} - M\}$, such that {$\tr \left(M \tilde{\sigma_0}\right) = T^{(1)}_{0j}$,  $\tr \left(M \tilde{\sigma_1}\right) = T^{(1)}_{1j}$, while $\tr \left((\mathbb{I} - M)\tilde{\sigma_0}\right) = T^{(-1)}_{0j}$,  $\tr \left((\mathbb{I} - M) \tilde{\sigma_1}\right) = T^{(-1)}_{1j}$.}

The convexity of quantum states allows us to express $\tilde{\sigma_1}$ as a convex combination of pure states $\tilde{\sigma_1} = \sum_i \lambda_i \ketbra{\psi_i}{\psi_i}$. Moreover, the subset of quantum states $\rho$ satisfying the linear constraint $\tr\left(M \rho\right) = k$ is also a convex set. Thus, we can decompose $\sigma_1$ into some $\{\lambda_i, \ket{\psi_i}\}_i$ such that $\tr\left(M \ketbra{\psi_i}{\psi_i} \right) = \tr\left(M \tilde{\sigma}_1\right)$ for each $i$ individually.

Letting $p_0$ and $p_1$ be the probabilities of classical causal states $s_0$ and $s_1$ respectively, we write the stationary state as $\tilde{\rho} = p_0 \tilde{\sigma}_0 + p_1 \tilde{\sigma}_1$, and explicitly factorise:
\begin{align}
\tilde{\rho} & = p_0 \tilde{\sigma}_0 + p_1 \sum_i \lambda_i \ketbra{\psi_i}{\psi_i} \nonumber \\
& = \sum_i \lambda_i \left( p_0 \tilde{\sigma}_0 + p_1 \ketbra{\psi_i}{\psi_i} \right).
\end{align}
Hence, from concavity of entropy,
\begin{align}
\Ent{\tilde{\rho}} &\geq  \sum_i \lambda_i \Ent{ p_0 \tilde{\sigma}_0 + p_1 \ketbra{\psi_i}{\psi_i}} \nonumber \\
& \geq
\min_i \{ \Ent{ p_0 \tilde{\sigma}_0 + p_1 \ketbra{\psi_i}{\psi_i}} \}.
\end{align}

Let $\ket{\psi_{\rm min}}$ be the choice of state that minimizes the above expression.
Any model that encodes onto the pair of states $\tilde{\sigma}_0$ and $\ket{\psi_{\rm min}}$ is guaranteed to have complexity equal or lower than $C(\mathcal{Q}')$.
Moreover since $\{M, \mathbb{I} - M\}$ acting on $\tilde{\sigma}_0$ and $\ket{\psi_{\rm min}}$ produces output statistics with fidelity $F_{\max} = F(P(\future{X} | s_0), P(\future{X}| s_1))$, by monotonicity of fidelity under contractive maps we have $F( \tilde{\sigma}_0, \ketbra{\psi_{\rm min}}{\psi_{\rm min}}) \le F_{max}$.

Exactly the same argument can be made to find some substitution $\ket{\phi_{\rm min}}$ for $\tilde{\sigma}_0$, generating a new encoding which is guaranteed to satisfy the maximum fidelity, and to result in equal or lower complexity.
This construction shows that no valid mixed state encoding can exceed all pure states encodings.
Hence, if we can find a pure state model such that the two internal states saturate the fidelity bound, then it will have lower entropy than any mixed state model.
\end{proof}
\end{lemma}

\end{document}